%% file: main.tex
\pdfoutput=1
\documentclass[9pt,conference]{IEEEtran}
\IEEEoverridecommandlockouts

\usepackage{cite}
\usepackage{amsmath,amssymb,amsfonts}
\usepackage{graphicx}
\usepackage{multirow}
\usepackage{makecell}
\usepackage{array}
\usepackage{booktabs}
\usepackage{tabularx}
\usepackage{url}
\usepackage{placeins}
\usepackage{xcolor}
\usepackage[colorlinks=true,linkcolor=black,citecolor=black,filecolor=black,urlcolor=blue]{hyperref}
\newcommand{\IEEEpapertextsize}{\fontsize{9pt}{11pt}\selectfont}

\renewcommand{\scriptsize}{\IEEEpapertextsize}

\providecommand{\Description}[1]{}

\def\BibTeX{{\rm B\kern-.05em{\sc i\kern-.025em b}\kern-.08em
    T\kern-.1667em\lower.7ex\hbox{E}\kern-.125emX}}

\begin{document}
\bstctlcite{IEEEexample:BSTcontrol}

\title{TokenJoule: Energy Modeling for Distributed Training and Inference of LLMs}

\author{\IEEEauthorblockN{George Karfakis\IEEEauthorrefmark{1},
Lime Yao\IEEEauthorrefmark{1},
Joyjit Kundu\IEEEauthorrefmark{2},
Debjyoti Bhattacharjee\IEEEauthorrefmark{2},
Puneet Gupta\IEEEauthorrefmark{1}}
\IEEEauthorblockA{\IEEEauthorrefmark{1}University of California, Los Angeles, USA\qquad
\IEEEauthorrefmark{2}imec, Belgium}}

\maketitle

\begin{abstract}
\input{sections/abstract.tex}
\end{abstract}


\input{sections/introduction.tex}
\input{sections/background.tex}
\input{sections/method.tex}

\input{sections/evaluation.tex}

\input{sections/summary.tex}

\clearpage
\bibliographystyle{IEEEtran}
\bibliography{ref}

\end{document}

%% file: sections/abstract.tex
As large language models (LLMs) scale, energy dictates operating cost, cooling, and cluster capacity across training and inference. Early hardware design-space exploration (DSE) requires predicting energy before silicon exists, yet performance models miss hidden component work while prior energy predictors calibrate to deployed GPUs. We present TokenJoule, an open-source analytical tool that derives device kernel sequences, tiled compute, shared-memory staging, cache residency, and collective traffic, pricing activity directly with component energies. With zero workload fitting, TokenJoule achieves 3.3--6.6\% activity error (151 A100/H100 runs) and 3.0--6.8\% A100 energy error, guiding parallelism, memory packaging, and 3D-DRAM chip co-design.

%% file: sections/introduction.tex
\section{Introduction}
\label{sec:introduction}

As large language models (LLMs) scale in size and deployment, energy has become a primary constraint on training and serving systems. It dictates electricity cost, cooling and power-delivery design, and system scale. To guide early design-space exploration (DSE), architects need models that predict how candidate hardware parameters and execution strategies jointly affect energy.

\begin{figure}[t]
\centering
\includegraphics[width=\columnwidth]{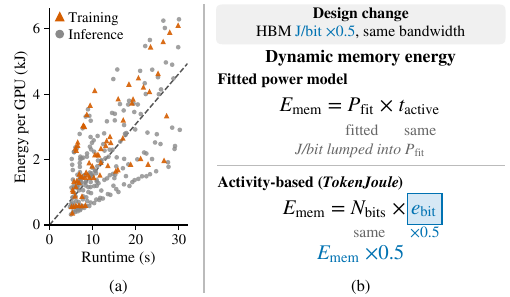}
\caption{(a) Measured energy per GPU vs. runtime for A100 LLM training and inference runs (dashed line: single average power). (b) Halving HBM energy per bit at fixed bandwidth is not parameterized in fitted power models ($P_{\mathrm{fit}} \times t_{\mathrm{active}}$, regressed on whole-GPU power), whereas TokenJoule halves $e_{\mathrm{bit}}$ directly.}
\label{fig:intro-scatter}
\end{figure}

Existing performance models for early LLM system exploration~\cite{calculon2023,deepflow_2024,llmcompass2024,kundu2024performance,rapidllm2026} are ill-suited to this task: they predict runtime from the bottleneck, whereas energy depends on the total work across compute and the memory hierarchy. Off-chip traffic hidden behind a compute-bound matrix multiply, padded arithmetic in narrow decode matrices, and repeated on-chip operand reads can all draw energy without affecting runtime~\cite{energaizer2026}. Measured A100 runs whose runtimes match within 10\% differ by up to 6.6$\times$ in energy per GPU because their total work differs (Figure~\ref{fig:intro-scatter}a). No single average power maps runtime to energy, so energy must be computed from activity.

Evaluating unbuilt hardware further requires pricing this activity without empirical fitting, since candidate designs have no silicon measurements to calibrate against. Cycle-level GPU power simulators~\cite{leng2013gpuwattch,micro-21-kan} cannot scale to distributed LLM execution, while analytical accelerator estimators~\cite{accelergy2019,timeloop2019} target per-layer spatial dataflows and do not derive multi-kernel GPU or collective activity. Recent LLM energy predictors instead rely on measured-GPU calibration: learned models~\cite{wattgpu2026,piep2025} regress measured energy on workload features, while fitted analytical models either calibrate coefficients to coarse utilization or traffic terms~\cite{sweetspot2026,enskat2026mfu} or map isolated-kernel activity to module busy times scaled by regressed power coefficients~\cite{energaizer2026,energylens2026}. Fitting onto whole-GPU power lumps per-access energy with component speed and bus width, leaving knobs such as memory joules per bit (Figure~\ref{fig:intro-scatter}b) or cache size unparameterized.

Hardware DSE must also cover both training and inference. Deployed accelerators often serve both, which stress compute, memory, and interconnects differently: training is shaped by capacity limits, activation recomputation, and gradient collectives, whereas decode is governed by small-batch underutilization and memory accesses. A design change that saves energy in one regime can be ineffective in the other, yet existing LLM predictors cover only one (Table~\ref{tab:related_work}).

To support calibration-free hardware DSE across both regimes, we present TokenJoule, an analytical energy model for distributed LLM training and inference. Given a workload, parallelization strategy, and hardware configuration, TokenJoule derives accurate component-level activity across the full multi-kernel execution, tracking compute units and the memory hierarchy as hardware parameters and mappings change. Pricing these predicted accesses directly with component energy constants and static power lets architects vary speeds, capacities, and per-access energies independently.

\begin{table}[t]
\centering
\caption{Comparison of energy-modeling capabilities.}
\label{tab:related_work}
\scriptsize
\setlength{\tabcolsep}{3pt}
\renewcommand{\arraystretch}{1.1}

\begin{tabular}{lccccl}
\toprule

\textbf{Work} &
\textbf{Inf.} &
\textbf{Tr.} &
\textbf{Dist.} &
\makecell{\textbf{Energy}\\\textbf{DSE}}&
\textbf{Approach}\\

\midrule

SweetSpot~\cite{sweetspot2026} &
$\checkmark$ & $\times$ & $\times$ &$\times$  &
Analytical + fit \\

EnergAIzer~\cite{energaizer2026} &
$\checkmark$ & $\times$ & $\times$ & $\circ$ &
Analytical + fit \\

EnergyLens~\cite{energylens2026} &
$\checkmark$ & $\times$ & $\checkmark$ &$\times$  &
Analytical + fit \\

Enskat \& Wiesner~\cite{enskat2026mfu} &
$\times$ & $\checkmark$ & $\times$ &$\times$  &
Analytical + fit \\

WattGPU~\cite{wattgpu2026} &
$\checkmark$ & $\times$ & $\times$ &$\times$  &
Learned \\

PIE-P~\cite{piep2025} &
$\checkmark$ & $\times$ & $\checkmark$ & $\times$ &
Learned\\
\midrule
\textbf{TokenJoule} &
$\checkmark$ & $\checkmark$ & $\checkmark$ &$\checkmark$  &
\textbf{Analytical} \\
\bottomrule
\end{tabular}
\vspace{3pt}
\begin{minipage}{\columnwidth}
\scriptsize
Inf.: Inference, Tr.: Training, Dist.: Distributed. $\circ$: Scales cores/BW/clock at fixed energy efficiency.
\end{minipage}

\end{table}

Compared with prior LLM energy predictors (Table~\ref{tab:related_work}), our contributions are:
\begin{itemize}
\item \textbf{Full-stack activity formulation:} Deriving per-device compute, memory, and collective activity across distributed LLM training and inference without profiling (Section~\ref{sec:method}).
\item \textbf{Component and system validation:} Validating activity within 3.3\%/6.6\% mean absolute percentage error (MAPE) across 151 A100/H100 runs against Nsight Compute (NCU) hardware counters, step energy within 3.4\%/4.4\% (single-GPU) and 6.8\%/3.0\% (multi-GPU inference/training) with 4.3--8.4\% step-time MAPE, and external inference energy within 9.5\% versus 16.1\% for EnergyLens (Section~\ref{sec:results}).
\item \textbf{Cross-stack hardware DSE:} Demonstrating parallelism co-search, memory packaging trade-offs, and power-capped 3D-DRAM floorplan exploration across training and inference (Section~\ref{sec:codesign}).
\end{itemize}

%% file: sections/background.tex
\section{Background}
\label{sec:background}

Distributed LLM training and inference partition the workload across devices via data, tensor, pipeline, and expert parallelism (DP, TP, PP, and EP) into per-device operations interleaved with collective communication. Although training, prefill, and decode stress the system differently, the software runtime lowers each device's operations into a sequence of kernels that all use the same hardware units. Kernels compute on matrix and vector units and move data through three memory levels: per-core shared memory (SMEM), a shared L2 cache, and off-chip DRAM (such as HBM on GPUs). Collective communication transfers data across inter-device links.

Energy per operation varies by more than an order of magnitude across these components. On modern GPUs, a vector FLOP costs roughly 18$\times$ a matrix FLOP, and DRAM access energy is roughly 3$\times$ that of L2 and 8$\times$ that of SMEM~\cite{antepara2025energy}. Predicting dynamic energy therefore requires tracking activity $N_i$ (FLOPs or bytes) at each compute unit, memory level, and network link and weighting it by the corresponding energy cost $e_i$. Adding static power $P_{\mathrm{static}}$ over step time $T$ yields total step energy:
\[
E = \sum_{i} N_i \, e_i + P_{\mathrm{static}} \, T.
\]

%% file: sections/method.tex
\section{TokenJoule Methodology}
\label{sec:method}
\input{sections/methodology.tex}

%% file: sections/methodology.tex
\begin{figure}[t]
\centering
\includegraphics[width=\columnwidth]{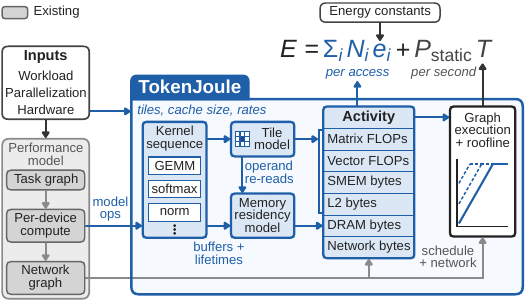}
\caption{Overview of TokenJoule: per-device kernel expansion, tiled execution modeling, and a step-level memory residency model predict hardware activity, which combines with independent component energy constants and static power.}
\Description{Separate platform characterization and analytical workload execution feed energy prediction.}
\label{fig:framework-overview}
\end{figure}

\subsection{Overview}
\label{sec:method-overview}
\label{sec:method-sequence}

Given a workload, parallelization strategy, and target hardware, an existing distributed performance model (we use RAPID-LLM~\cite{rapidllm2026}) supplies the per-device task graph of logical compute and collective operations alongside collective network bytes. These task-level graphs capture inter-device transfers, but do not resolve on-device hardware activity. From each device's operations, TokenJoule independently derives the on-device terms of $N_i$: issued matrix and vector FLOPs, and SMEM, L2, and off-chip DRAM bytes.

Following Figure~\ref{fig:framework-overview}, TokenJoule reconstructs each device's kernel sequence with tensor buffer sizes and lifetimes from the framework's operator-fusion, activation-recomputation, and master-weight precision semantics. Alongside matrix multiplications (GEMMs, including mixture-of-experts (MoE) layers), attention, and non-matrix kernels, this sequence tracks weights, key-value (KV) cache state in prefill and decode, and activations, gradients, and optimizer states in training. From this sequence, TokenJoule models kernel compute and SMEM (Section~\ref{sec:method-tile}), step-level L2 and DRAM traffic (Section~\ref{sec:method-l2}), collective communication (Section~\ref{sec:method-network}), and step time and energy pricing (Section~\ref{sec:method-pricing}).

\begin{figure}[t]
\centering
\includegraphics[width=\columnwidth]{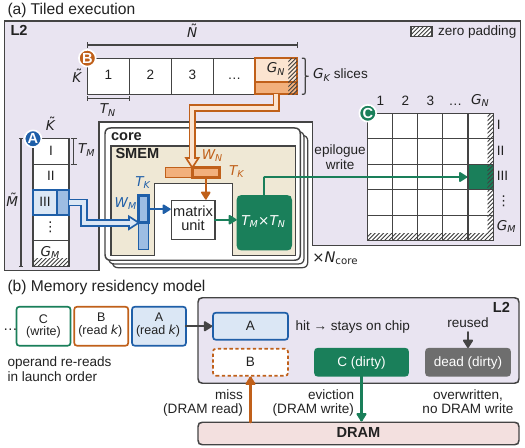}
\caption{(a) Across $N_{\mathrm{core}}$ cores, each $T_M \times T_N$ output tile of the zero-padded $G_M \times G_N$ grid stages $T_K$ slices in SMEM, feeds $(W_M, W_N)$ sub-tiles to the matrix unit, and writes the accumulated $C$ tile to L2 once, re-reading row blocks of $A$ and column blocks of $B$ across the grid. (b) In the memory residency model, operand re-reads and writes arrive in launch order at the fixed-size L2: hits stay on chip, misses and dirty evictions become off-chip DRAM bytes, and reused dead allocations are overwritten without DRAM writes.}
\label{fig:method-mechanism}
\end{figure}

\subsection{Kernel Compute and SMEM}
\label{sec:method-tile}
Mapping each kernel in the sequence onto hardware begins with tiling, as tile dimensions determine issued FLOPs (including zero-padding), intra-tile SMEM staging, and inter-tile L2 re-reads, each a major driver of dynamic energy.

\begin{figure}[t]
\centering
\includegraphics[width=\columnwidth]{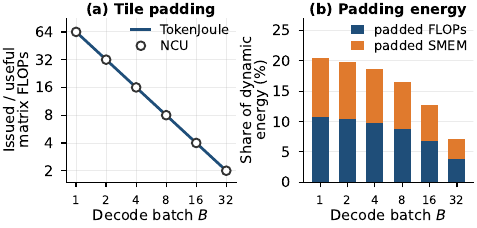}
\caption{Llama-3.1-8B decode on A100 across small batch sizes $B = 1\text{--}32$: (a) issued-to-useful matrix FLOP ratio from Nsight Compute (NCU) hardware counters and TokenJoule, and (b) share of step dynamic energy from padded matrix FLOPs and padded SMEM traffic.}
\label{fig:padding-motivator}
\end{figure}

\noindent\textbf{Matrix kernels.} Because tiled datapaths execute fixed-width blocks, an $(M, N, K)$ GEMM $C = AB$ decomposes into feasible $(T_M, T_N, T_K)$ tiles within matrix-unit, occupancy, and SMEM limits, slicing large dimensions and zero-padding narrow ones (Figure~\ref{fig:method-mechanism}a). TokenJoule analytically selects the feasible tile that maximizes useful (unpadded) throughput. Forming a grid of $G_d = \lceil d/T_d \rceil$ tiles ($d \in \{M, N, K\}$) pads each dimension to $\tilde{d} = G_d T_d$ and issues $F = 2\tilde{M}\tilde{N}\tilde{K}$ matrix FLOPs, pricing useful ($2MNK$) and zero-padded ($F - 2MNK$) operations separately for lower zero-input switching~\cite{antepara2025energy}. Staging tiles in SMEM across register $(W_M, W_N)$ sub-tiles and writing $C$ once generates $\frac{F}{2}(\frac{1}{W_M} + \frac{1}{W_N})$ element reads and $MN$ element writes in SMEM, scaled by precision bytes. On A100 Llama-3.1-8B decode (Figure~\ref{fig:padding-motivator}), narrow batches ($M < T_M$) inflate issued matrix FLOPs up to 64$\times$ with matching SMEM traffic behind HBM-bound runtime, reaching 20.5\% of dynamic energy.

\noindent\textbf{Attention and vector kernels.} Fused attention chains tiled GEMMs with vector-unit online softmax in SMEM via FlashAttention~\cite{dao2023flashattention2} without materializing intermediate score buffers. Standalone elementwise, normalization, optimizer, and collective kernels (plus occasional library-dispatched matrix-vector decode kernels) execute on vector units without SMEM staging, issuing vector FLOPs from logical shapes.

\subsection{L2 and DRAM Traffic}
\label{sec:method-l2}
Vector kernels stream each tensor buffer once through L2, whereas tiled matrix kernels re-read inputs across tiles that stage operands independently in SMEM. Across the $G_M \times G_N$ grid, cores fetch $A$ from L2 $G_N$ times, fetch $B$ $G_M$ times, and write $C$ once (Figure~\ref{fig:method-mechanism}a), with analogous passes in fused attention. Summing these passes and writes across the kernel sequence gives total L2 bytes.

Off-chip DRAM bytes come from L2 misses and dirty evictions from the fixed-size cache, both within a multi-pass GEMM and across consecutive kernels (Figure~\ref{fig:method-mechanism}b). Because LLM kernels sweep contiguous buffers in structured passes, TokenJoule captures both effects without tracking individual cache lines using a memory residency model over Section~\ref{sec:method-overview}'s buffer sizes and lifetimes. Walking the sequence in launch order, the model tracks resident buffer ranges in L2 as new passes arrive and displace older data. Compact multi-pass inputs ($A$) and activations passed between unfused kernels hit on chip when they fit, whereas large weight tensors ($B$) exceed L2, evict earlier data across sweeps, and read from DRAM. Kernel outputs ($C$) remain in L2 for consumers: dirty evictions write back to DRAM, whereas freed buffers overwritten in place before eviction avoid DRAM writes.

Table~\ref{tab:l2-ablation} motivates both stages across all 99 A100 runs: omitting tiling misses L2 tile re-reads (47.8\% vs.\ 3.0\% error), whereas under tiling, zero L2 (263\% HBM error), per-kernel ideal reuse~\cite{energaizer2026} (15.2\%), and infinite L2 (43.5\%) mispredict HBM traffic, and cold per-kernel finite-L2 error (10.6\%) halves once L2 state persists across kernels in TokenJoule (5.3\%).

\begin{table}[t]
\centering
\scriptsize
\caption{L2 and HBM byte MAPE against A100 NCU hardware counters across the 99 A100 runs (Section~\ref{sec:method-validation}).}
\label{tab:l2-ablation}
\setlength{\tabcolsep}{4.5pt}
\renewcommand{\arraystretch}{1.08}
\begin{tabular}{@{}llcc@{}}
\toprule
\textbf{Tiling} & \textbf{L2 Cache Model} & \textbf{L2 MAPE} & \textbf{HBM MAPE} \\
\midrule
Untiled & None (size 0) & 47.8\% & 15.2\% \\
Tiled & None (size 0) & 3.0\% & 263\% \\
Tiled & Per-kernel ideal~\cite{energaizer2026} & 3.0\% & 15.2\% \\
Tiled & Infinite & 3.0\% & 43.5\% \\
Tiled & Finite, cold per kernel & 3.0\% & 10.6\% \\
\textbf{Tiled} & \textbf{Finite, persistent (TokenJoule)} & \textbf{3.0\%} & \textbf{5.3\%} \\
\bottomrule
\end{tabular}
\end{table}

\subsection{Collective Communication}
\label{sec:method-network}
Distributed parallelisms interleave compute kernels with tensor, expert, pipeline, and data-parallel collectives that draw energy both on device and across interconnect links. On each device, reduction collectives (AllReduce and ReduceScatter) contribute vector-priced reduction FLOPs, while analytical collective formulas capture buffer traffic through L2 and DRAM, displacing cached compute tensors in the residency model. Across devices, the distributed task graph applies the collective algorithm across each network dimension to determine transmitted bytes priced at that link's energy per byte. Parallelism also feeds back into Section~\ref{sec:method-tile}: tensor and expert sharding narrow per-device matrix widths and token rows, altering tile padding and cache reuse.

\subsection{Step Time and Energy Pricing}
\label{sec:method-pricing}
Because static energy $P_{\mathrm{static}} T$ acts over the full step across devices, TokenJoule predicts each kernel's duration from its tiled GEMM or multi-level roofline model across compute, SMEM, L2, and DRAM, bounded below by a kernel launch floor (Figure~\ref{fig:framework-overview}). Task-graph scheduling then combines per-device kernels with collective transfers, capturing compute-communication overlap and scheduled pipeline bubbles in the step completion time $T$.

Substituting predicted $(N_i, T)$ into Section~\ref{sec:background}'s energy equation with component parameters $(e_i, P_{\mathrm{static}})$ yields step energy $E$, scaling on-device dynamic access energies with squared operating voltage when clock frequency varies. Deployed accelerators obtain $(e_i, P_{\mathrm{static}})$ from published microbenchmarks and idle power (Section~\ref{sec:results}), whereas unbuilt designs specify candidate architectural and circuit parameters directly (Section~\ref{sec:codesign}).

%% file: sections/evaluation.tex
\section{Validation}
\label{sec:results}
\label{sec:local-validation}
\label{sec:network-validation}
\label{sec:distributed}
\label{sec:inference-validation}
\label{sec:training-validation}

\subsection{Activity Validation}
\label{sec:method-validation}
\begin{figure*}[t]
\centering
\includegraphics[width=0.96\textwidth]{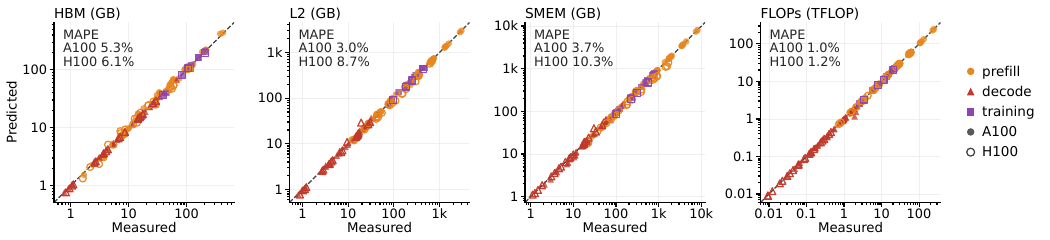}
\caption{Predicted versus measured hardware activity across 151 full-step runs on A100 (99 dense and MoE runs) and H100 (52 runs).}
\Description{Predicted-to-measured activity across memory levels and matrix operations on A100 and H100.}
\label{fig:activity-validation}
\label{fig:counter-energy-reconstruction}
\label{fig:h100-counts}
\end{figure*}

Validating $N_i$ directly before applying energy prices guards against cross-component error cancellation across architectures. Using Nsight Compute (NCU)~\cite{ncu} hardware counters under non-flushing full-step replay on NVIDIA A100 and H100 GPUs (80~GB), we measure 151 runs (99 A100, 52 H100) across vLLM prefill, vLLM decode, and training on dense and MoE Llama~\cite{touvron2023llama2,grattafiori2024llama3}, SmolLM2~\cite{allal2025smollm2}, and Qwen~\cite{yang2024qwen25,yang2025qwen3} models (0.36B--30B). Figure~\ref{fig:activity-validation} compares predicted and measured HBM, L2, and SMEM bytes and issued matrix FLOPs, achieving 3.3\% overall MAPE on A100 and 6.6\% on H100.

\subsection{Experimental Methodology}
\label{sec:eval-setup}

\begin{table}[t]
  \centering
  \scriptsize
  \caption{A100 (80~GB) hardware and energy inputs (1410~MHz reference~\cite{antepara2025energy}, scaled by $V^2$ at run clock~\cite{schoonhoven2022going}).}
  \label{tab:a100-model-inputs}
  \setlength{\tabcolsep}{4pt}
  \renewcommand{\arraystretch}{1.08}
  \begin{tabular}{@{}lr@{}}
    \toprule
    \textbf{Parameter} & \textbf{Value} \\
    \midrule
    \multicolumn{2}{@{}l}{\textit{Datasheet specifications}~\cite{nvidia2021a100specs}} \\
    HBM bandwidth (PCIe / SXM4) & 1.94 / 2.04~TB/s \\
    SMEM per SM / L2 capacity & 164~KB / 40~MB \\
    \midrule
    \multicolumn{2}{@{}l}{\textit{GPU microbenchmarks (1410~MHz)}~\cite{antepara2025energy}} \\
    SMEM / L2 bandwidth & 14.8 / 4.3~TB/s \\
    SMEM / L2 / HBM latency~\cite{luo2025hopper} & 29 / 203 / 566~cycles \\
    FP16 matrix / padded energy & 0.70 / 0.37~pJ/FLOP \\
    FP16 vector energy & 12.97~pJ/FLOP \\
    SMEM (L1 SRAM) / L2 energy & 12.72 / 37.68~pJ/B \\
    HBM access energy & 104.88~pJ/B \\
    \midrule
    \multicolumn{2}{@{}l}{\textit{Interconnect and runtime}} \\
    Inter-GPU link energy~\cite{dally2020gtcchina} & 64~pJ/B (8~pJ/b) \\
    Kernel launch floor~\cite{vellaisamy2026taxbreak} & 5~$\mu$s \\
    \bottomrule
  \end{tabular}
\end{table}

We validate end-to-end step energy on NVIDIA A100 GPUs using published component-level hardware and energy constants (Table~\ref{tab:a100-model-inputs}). Applying these parameters with zero LLM workload fitting, we evaluate a single A100 PCIe 80~GB GPU and an 8$\times$ A100 SXM4 80~GB NVLink node. Inference uses vLLM (prefill and decode) and training uses HuggingFace PyTorch and Megatron-Core~\cite{eff_megatron}. Single-GPU runs lock the 1065~MHz base clock to avoid boost and power-capping transients, reading NVML~\cite{nvml} cumulative energy over synchronized windows of repeated steps after warmup while replaying prefill and decode separately and removing inter-step harness idle energy. For multi-GPU runs, measured step energy sums NVML readings across GPUs, including on-package NVLink interface power while excluding external NVSwitch chips. Following Antepara et al.~\cite{antepara2025energy}, $P_{\mathrm{static}}$ is unloaded GPU idle power (using their published A100 value for external EnergyLens runs~\cite{energylens2026}).

\subsection{End-to-End Energy Validation}
\label{sec:energy-validation}

\begin{figure}[t]
\centering
\includegraphics[width=0.90\columnwidth]{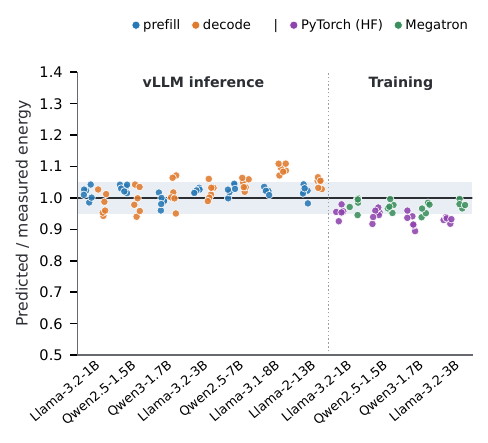}
\caption{Single-GPU predicted-to-measured energy across 127 A100 runs (shaded band marks $\pm$5\%): 84 vLLM prefill and decode runs (3.4\% MAPE) and 43 training runs (4.4\% MAPE: 23 HuggingFace PyTorch, 20 Megatron).}
\Description{Single-GPU predicted-to-measured energy ratios.}
\label{fig:validation-summary}
\label{fig:local-qwen-energy}
\label{fig:local-llama-prefill-energy}
\label{fig:local-llama-decode-energy}
\end{figure}

\begin{figure}[t]
\centering
\includegraphics[width=0.90\columnwidth]{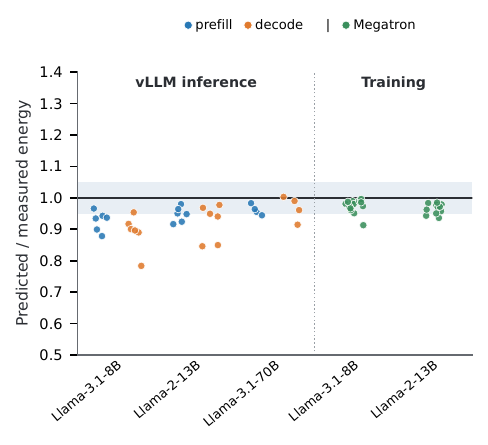}
\caption{Distributed 2--8$\times$ A100 SXM predicted-to-measured energy across 59 runs (shaded band marks $\pm$5\%): 32 vLLM tensor-parallel (TP = 2, 4, 8) prefill and decode runs (6.8\% MAPE) and 27 Megatron training runs (3.0\% MAPE).}
\label{fig:multigpu-energy}
\end{figure}

\begin{figure}[t]
\centering
\includegraphics[width=0.88\columnwidth]{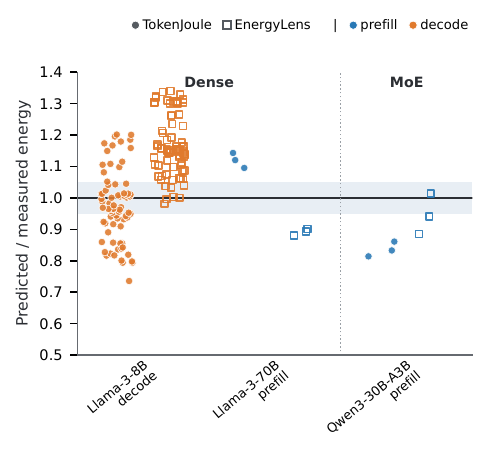}
\caption{Predicted-to-measured inference energy for TokenJoule (zero workload fitting, 9.5\% MAPE) versus calibrated EnergyLens~\cite{energylens2026} (16.1\% MAPE) across 81 dense and MoE runs (shaded band marks $\pm$5\%).}
\Description{TokenJoule and EnergyLens energy comparison.}
\label{fig:energylens-comparison}
\end{figure}

\noindent\textbf{Single-GPU inference and training.} Across 127 single-GPU runs sweeping batch sizes and sequence lengths on models from 1B to 13B parameters (Figure~\ref{fig:validation-summary}), TokenJoule achieves 3.4\% energy MAPE (8.4\% step-time MAPE) on vLLM inference and 4.4\% energy MAPE (5.8\% step-time MAPE) on training. Dense prefill tracks measured energy closely because large tiles amortize runtime overheads, while prediction errors vary more across decode workloads as small-batch kernels become sensitive to launch overheads and narrow-tile dispatch. In training, backward gradients, precision casts, and optimizer updates introduce auxiliary kernels and fragmented launches that cause slight underprediction on eager HuggingFace PyTorch, whereas heavier fusion in Megatron keeps ratios near unity.

\noindent\textbf{Distributed inference and training.} Across 59 runs sweeping TP, DP, and PP configurations on an 8$\times$ A100 SXM node up to 70B parameters (Figure~\ref{fig:multigpu-energy}), TokenJoule achieves 6.8\% energy MAPE (5.6\% step-time MAPE) on tensor-parallel vLLM inference and 3.0\% energy MAPE (4.3\% step-time MAPE) on distributed Megatron training. In tensor-parallel decode, sharding shrinks already tiny kernels while adding two AllReduce synchronizations per layer, so unpredictable host launch slack and barrier jitter add measured static energy. Distributed prefill and Megatron training amortize these synchronization costs over much larger tensors and coarse gradient collectives, keeping predictions within ${\sim}\pm$10\%.

\noindent\textbf{Comparison with EnergyLens.} We compare TokenJoule predictions against published measurements across the 81 single- and multi-GPU inference runs with consistent per-request metrics in EnergyLens~\cite{energylens2026} (Figure~\ref{fig:energylens-comparison}), spanning dense Llama-3-8B decode and Llama-3-70B prefill alongside MoE Qwen3-30B-A3B prefill. With zero target-system or workload fitting, TokenJoule achieves 9.5\% MAPE versus 16.1\% for EnergyLens's calibrated predictor, supporting calibration-free exploration of unbuilt hardware in Section~\ref{sec:codesign}.

\input{sections/codesign.tex}

%% file: sections/codesign.tex
\section{Design-Space Exploration Case Studies}
\label{sec:codesign}

The following case studies illustrate early design questions TokenJoule can answer before silicon exists. Because TokenJoule derives component activity without measured-workload fitting, each study starts from our validated A100 baseline (Table~\ref{tab:a100-model-inputs}) and varies capacity, rates, packaging, or circuit energies across candidate designs.

\label{sec:hardware-whatifs}

\subsection{Hardware Pathfinding with Configuration Co-Search}
\label{sec:cs1-pathfinding}

\begin{figure}[t]
\centering
\includegraphics[width=0.94\columnwidth]{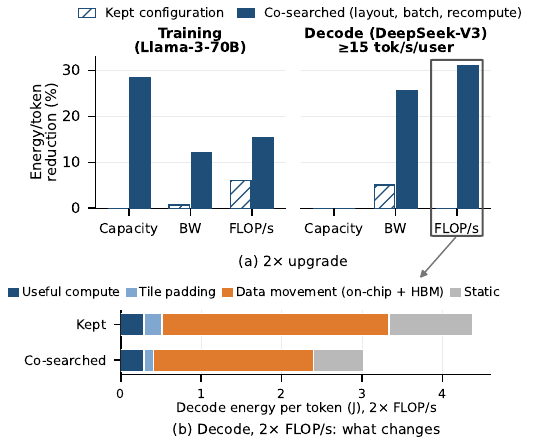}
\caption{(a) Energy-per-token reduction from 2$\times$ DRAM capacity, HBM bandwidth (BW), or matrix FLOP/s over A100-80GB under kept versus co-searched configurations. (b) Per-token decode energy breakdown under 2$\times$ FLOP/s.}
\label{fig:hardware-reselection}
\end{figure}

Hardware upgrades are often evaluated under a fixed software deployment, ignoring how new resources unlock more efficient parallelisms, batch sizes, or recomputation schedules. Starting from an energy-optimal baseline configuration on A100-80GB clusters for 64-GPU Llama-3-70B training (8K sequence~\cite{nemo2507}) and 144-GPU DeepSeek-V3 decode (16K context, $\ge$15~tokens/s/user~\cite{deepseekv3_2024}), we double DRAM capacity, HBM bandwidth, or matrix FLOP/s one at a time and compare keeping the baseline configuration against co-searching the deployment (Figure~\ref{fig:hardware-reselection}).

With the configurations kept, no upgrade saves more than 6\%: added capacity sits idle, and faster bandwidth or compute finishes the same work sooner, saving only static energy (Figure~\ref{fig:hardware-reselection}a). Co-searching turns the same upgrades into savings of up to 31\% and changes which one wins. In training, doubling the capacity saves the most (28\%) by removing activation recomputation and reducing tensor-parallel sharding. In decode, capacity already suffices, but latency limits batching, so 2$\times$ FLOP/s overtakes 2$\times$ bandwidth (31\% versus 26\%) by permitting larger batches that amortize data movement, narrow-tile padding (Section~\ref{sec:method-tile}), and static energy while useful compute per token stays unchanged (Figure~\ref{fig:hardware-reselection}b). Upgrades thus pay off mainly through the deployments they enable, which TokenJoule prices component by component in both regimes.

\subsection{Memory Packaging for Serving}
\label{sec:inference-interactivity}

\begin{table}[t]
\centering
\scriptsize
\caption{Memory systems at 1.555~TB/s bandwidth (total includes 8.5~pJ/b of on-chip overhead~\cite{antepara2025energy}).}
\label{tab:cs2-memory-inputs}
\setlength{\tabcolsep}{4.5pt}
\renewcommand{\arraystretch}{1.1}
\begin{tabular}{@{}lrl@{}}
\toprule
\textbf{Memory} & \textbf{Cap.} & \textbf{Total (DRAM + Link/Switch)} \\
\midrule
A100 HBM & 40~GB & 13.1~pJ/b (4.6~\cite{antepara2025energy}) \\
Direct GDDR7 & 128~GB & 18.5~pJ/b (10.0~\cite{oconnor2017fgdram,cho2024gddr7}) \\
Disagg.\ LPDDR5X & 640~GB & 24.0~pJ/b (5.0~\cite{nvidia2023grace} + 10.5~\cite{park2023lr112g,credo2025weaver}) \\
\bottomrule
\end{tabular}
\end{table}

\begin{figure}[t]
\centering
\includegraphics[width=0.94\columnwidth]{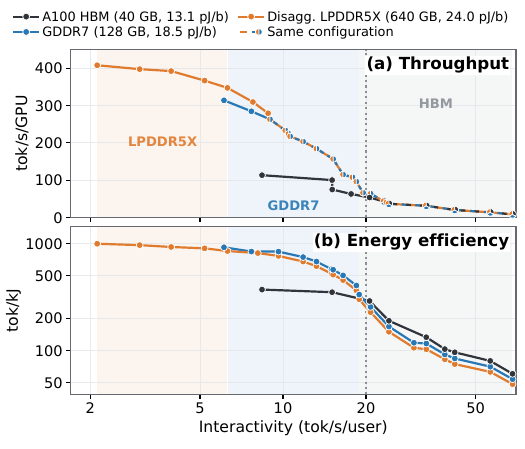}
\caption{Qwen3-235B-A22B serving (4K input, 1K output) Pareto frontiers across 1--16 GPUs (TP, DP, PP, and batch) at matched 1.555~TB/s bandwidth, shaded by the energy-efficiency winner across interactivity: (a) throughput and (b) energy efficiency.}
\label{fig:memory-efficiency-interactivity}
\end{figure}

LLM serving needs capacity for weights and multi-request KV caches so that larger batches can amortize weight reads and static power. On-package HBM caps that capacity: its wide, low-energy interfaces must line the compute die's shoreline on a reticle-limited interposer. Moving DRAM off the interposer removes the cap but costs more energy per bit. At matched 1.555~TB/s bandwidth and static power, we compare A100 HBM against board-mounted GDDR7 and SerDes-disaggregated LPDDR5X (Table~\ref{tab:cs2-memory-inputs}) for Qwen3-235B-A22B serving on 1--16 GPUs (Figure~\ref{fig:memory-efficiency-interactivity}).

The winner depends on interactivity. Above 20~tokens/s/user, batches are small, so all three memories run the same configuration at the same speed. Only energy per bit separates them: HBM delivers 12--15\% more tokens per joule than GDDR7 and 25--29\% more than LPDDR5X. Below 20~tokens/s/user, larger memories admit larger batches: GDDR7 wins down to 6~tokens/s/user with up to 2.3$\times$ HBM's tokens per joule despite 1.4$\times$ higher energy per bit, and LPDDR5X wins below that, where smaller memories run out of capacity. TokenJoule finds this crossover for unbuilt GPU-memory pairings by pricing each access while capacity reshapes the deployment.

\subsection{Power-Capped 3D-DRAM Chip Co-Design}
\label{sec:cs3-3ddram}

\begin{figure}[t]
\centering
\includegraphics[width=0.92\columnwidth]{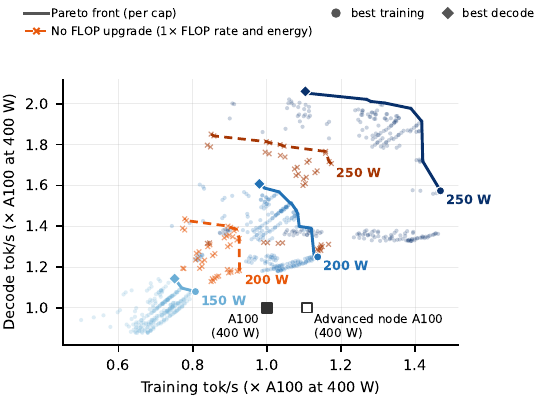}
\caption{DeepSeek-V2-Lite~\cite{deepseekv2_2024} 8-GPU training and decode throughput ($\times$ 400~W A100) across 826~mm$^2$ 3D-DRAM floorplans under 150, 200, and 250~W caps (250~W approximates A100 0.29~W/mm$^2$ 2.5D power density).}
\label{fig:cs3-3ddram}
\end{figure}

3D-stacked DRAM promises high bandwidth at low energy per bit~\cite{oconnor2017fgdram,bai2020sedram}, yet faces acute thermal constraints and design complexity~\cite{chen2025hbm3d}. Thermal coupling can degrade HBM performance in 2.5D packages~\cite{11113276}, and vertical stacking intensifies this limit. We use TokenJoule to showcase how architects can navigate these options, a co-design space missed by power-unaware floorplan models~\cite{llmcompass2024} and fitted single-regime predictors. Starting from an A100 die footprint (826~mm$^2$), we stack 3D DRAM under a 250~W cap matching the 400~W A100's 2.5D interposer W/mm$^2$ density (plus stricter 150 and 200~W caps), evaluating both a 2$\times$-faster, half-energy logic node and unscaled A100 compute across 8-GPU DeepSeek-V2-Lite~\cite{deepseekv2_2024} training and decode (Figure~\ref{fig:cs3-3ddram}).

We enumerate 726 candidate systems spanning SM allocation, L2 capacity (down to no L2), L2 bandwidth, and 4--6 DRAM stacks under area-scaled leakage and $V^2$ clock scaling~\cite{schoonhoven2022going}. Because each candidate evaluation is independent, the sweep is fully parallel and completes in tens of minutes.

Conventional 2.5D GPUs rely on a large L2 to shield costly off-die HBM accesses, whereas low-energy 3D DRAM prompts both regimes to shrink L2. At 250~W on the advanced node, best decode (48~SMs, 5~stacks, 2.06$\times$ A100) removes L2 completely given low KV reuse, while best training (88~SMs, 4~stacks, 1.47$\times$ A100) pares L2 to 0.25$\times$ yet retains it to prevent a 2.6$\times$ DRAM traffic jump, with balanced solutions within 5\% of both optima. Removing the FLOP upgrade (dashed fronts) confirms what drives each regime: 3D DRAM alone still achieves 1.85$\times$ decode at 250~W (1.44$\times$ at 200~W) with the same no-L2 floorplan, whereas training drops to 1.17$\times$ and 0.93$\times$ (near the 1.11$\times$ advanced-node 2.5D A100), showing 3D DRAM drives decode while training requires co-scaling logic and 3D DRAM.

%% file: sections/summary.tex
\section{Conclusion}
\label{sec:discussion}

We presented TokenJoule, an analytical energy model for distributed LLM training and inference that decouples full-stack hardware activity from component energy pricing. By resolving device kernel sequences, tiled compute and shared-memory staging, step-level L2 residency, and collective transfers without profiling, TokenJoule predicts component activity within 3.3--6.6\% across GPU generations and step energy within 3.0--6.8\% across single- and multi-GPU regimes. Grounded in physical component costs with zero workload fitting, TokenJoule enables cross-regime co-design of parallel mappings, memory packaging, and power-capped 3D-DRAM accelerators. TokenJoule will be released as a freely available open-source tool upon acceptance.

%% file: main.bbl
\begin{thebibliography}{10}
\providecommand{\url}[1]{#1}
\csname url@samestyle\endcsname
\providecommand{\newblock}{\relax}
\providecommand{\bibinfo}[2]{#2}
\providecommand{\BIBentrySTDinterwordspacing}{\spaceskip=0pt\relax}
\providecommand{\BIBentryALTinterwordstretchfactor}{4}
\providecommand{\BIBentryALTinterwordspacing}{\spaceskip=\fontdimen2\font plus
\BIBentryALTinterwordstretchfactor\fontdimen3\font minus \fontdimen4\font\relax}
\providecommand{\BIBforeignlanguage}[2]{{%
\expandafter\ifx\csname l@#1\endcsname\relax
\typeout{** WARNING: IEEEtran.bst: No hyphenation pattern has been}%
\typeout{** loaded for the language `#1'. Using the pattern for}%
\typeout{** the default language instead.}%
\else
\language=\csname l@#1\endcsname
\fi
#2}}
\providecommand{\BIBdecl}{\relax}
\BIBdecl
\renewcommand{\BIBentryALTinterwordstretchfactor}{4}

\bibitem{calculon2023}
M.~Isaev \emph{et~al.}, ``Calculon: a methodology and tool for high-level co-design of systems and large language models,'' in \emph{Proc. SC}.\hskip 1em plus 0.5em minus 0.4em\relax ACM, 2023, pp. 71:1--71:14.

\bibitem{deepflow_2024}
N.~Ardalani, S.~Pal, and P.~Gupta, ``{DeepFlow}: A cross-stack pathfinding framework for distributed {AI} systems,'' \emph{ACM Trans. Des. Autom. Electron. Syst.}, vol.~29, no.~2, Feb. 2024.

\bibitem{llmcompass2024}
H.~Zhang \emph{et~al.}, ``{LLMCompass}: Enabling efficient hardware design for large language model inference,'' in \emph{ACM/IEEE ISCA}, 2024, pp. 1080--1096.

\bibitem{kundu2024performance}
J.~Kundu \emph{et~al.}, ``Performance modeling and workload analysis of distributed large language model training and inference,'' in \emph{IEEE IISWC}.\hskip 1em plus 0.5em minus 0.4em\relax IEEE, 2024, pp. 57--67.

\bibitem{rapidllm2026}
G.~Karfakis \emph{et~al.}, ``{RAPID-LLM}: Resilience-aware performance analysis of infrastructure for distributed {LLM} training and inference,'' arXiv:2512.19606, 2026.

\bibitem{energaizer2026}
K.~Lee \emph{et~al.}, ``{EnergAIzer}: Fast and accurate {GPU} power estimation framework for {AI} workloads,'' in \emph{IEEE ISPASS}, 2026.

\bibitem{leng2013gpuwattch}
J.~Leng \emph{et~al.}, ``{GPUWattch}: Enabling energy optimizations in {GPGPUs},'' \emph{ACM SIGARCH Comput. Archit. News}, vol.~41, no.~3, pp. 487--498, 2013.

\bibitem{micro-21-kan}
V.~Kandiah \emph{et~al.}, ``{AccelWattch}: A power modeling framework for modern {GPUs},'' in \emph{IEEE/ACM MICRO}, 2021, p. 738–753.

\bibitem{accelergy2019}
Y.~N. Wu, J.~S. Emer, and V.~Sze, ``Accelergy: An architecture-level energy estimation methodology for accelerator designs,'' in \emph{IEEE/ACM ICCAD}, 2019.

\bibitem{timeloop2019}
A.~Parashar \emph{et~al.}, ``Timeloop: A systematic approach to {DNN} accelerator evaluation,'' in \emph{IEEE ISPASS}, 2019.

\bibitem{wattgpu2026}
M.~F. Argerich, J.~F{\"u}rst, and M.~Pati{\~n}o-Mart{\'i}nez, ``{WattGPU}: Predicting inference power and latency on unseen {GPUs} and {LLMs},'' IJCAI SustainAI Workshop, 2026.

\bibitem{piep2025}
A.~Dutt \emph{et~al.}, ``Fine-grained energy prediction for parallelized {LLM} inference with {PIE-P},'' arXiv:2512.12801, 2025.

\bibitem{sweetspot2026}
H.~P. Cavagna \emph{et~al.}, ``{SweetSpot}: An analytical model for predicting energy efficiency of {LLM} inference,'' in \emph{ACM/SPEC ICPE}, 2026.

\bibitem{enskat2026mfu}
N.~Enskat and P.~Wiesner, ``Evaluating {MFU} as a proxy for {GPU} power for energy-aware simulation of {LLM} training,'' 2026, mASCOTS 2026.

\bibitem{energylens2026}
Z.~Song \emph{et~al.}, ``{EnergyLens}: Predictive energy-aware exploration for multi-{GPU} {LLM} inference optimization,'' arXiv:2605.14249, 2026.

\bibitem{antepara2025energy}
O.~Antepara \emph{et~al.}, ``Benchmark-driven models for energy analysis and attribution of {GPU}-accelerated supercomputing,'' in \emph{Proc. SC}, 2025, pp. 888--904.

\bibitem{dao2023flashattention2}
T.~Dao, ``{FlashAttention-2}: Faster attention with better parallelism and work partitioning,'' in \emph{ICLR}, 2024.

\bibitem{ncu}
{NVIDIA Corporation}, \emph{{NVIDIA Nsight Compute} Documentation}, 2024.

\bibitem{touvron2023llama2}
H.~Touvron \emph{et~al.}, ``{Llama 2}: Open foundation and fine-tuned chat models,'' arXiv:2307.09288, 2023.

\bibitem{grattafiori2024llama3}
A.~Grattafiori \emph{et~al.}, ``The {Llama 3} herd of models,'' arXiv:2407.21783, 2024.

\bibitem{allal2025smollm2}
L.~Ben~Allal \emph{et~al.}, ``{SmolLM2}: When smol goes big -- data-centric training of a small language model,'' arXiv:2502.02737, 2025.

\bibitem{yang2024qwen25}
A.~Yang \emph{et~al.}, ``{Qwen2.5} technical report,'' arXiv:2412.15115, 2024.

\bibitem{yang2025qwen3}
------, ``{Qwen3} technical report,'' arXiv:2505.09388, 2025.

\bibitem{schoonhoven2022going}
R.~Schoonhoven \emph{et~al.}, ``Going green: Optimizing {GPUs} for energy efficiency through model-steered auto-tuning,'' in \emph{IEEE/ACM PMBS}, 2022, pp. 1--13.

\bibitem{nvidia2021a100specs}
{NVIDIA Corporation}, ``{NVIDIA A100 Tensor Core GPU}: {SXM4} and {PCIe} specifications,'' Jun. 2021.

\bibitem{luo2025hopper}
W.~Luo \emph{et~al.}, ``Dissecting the {NVIDIA Hopper} architecture through microbenchmarking and multiple level analysis,'' arXiv:2501.12084, 2025.

\bibitem{dally2020gtcchina}
W.~J. Dally, ``{GTC China 2020} keynote: Long reach signaling,'' NVIDIA GTC China, 2020.

\bibitem{vellaisamy2026taxbreak}
P.~Vellaisamy \emph{et~al.}, ``{TaxBreak}: Unmasking the hidden costs of {LLM} inference through overhead decomposition,'' in \emph{IEEE ISPASS}, 2026.

\bibitem{eff_megatron}
D.~Narayanan \emph{et~al.}, ``Efficient large-scale language model training on {GPU} clusters using {Megatron-LM},'' in \emph{Proc. SC}, 2021, pp. 1--15.

\bibitem{nvml}
{NVIDIA}, \emph{{NVML} {API} Reference Guide}, accessed October 6, 2026.

\bibitem{nemo2507}
{NVIDIA Corporation}, \emph{{NVIDIA NeMo} Framework 25.07 Pre-Training Performance Summary}, 2025.

\bibitem{deepseekv3_2024}
A.~Liu \emph{et~al.}, ``{DeepSeek-V3} technical report,'' arXiv:2412.19437, 2024.

\bibitem{oconnor2017fgdram}
M.~O'Connor \emph{et~al.}, ``Fine-grained {DRAM}: Energy-efficient {DRAM} for extreme bandwidth systems,'' in \emph{IEEE/ACM MICRO}, 2017, pp. 41--54.

\bibitem{cho2024gddr7}
S.-Y. Cho \emph{et~al.}, ``A {16Gb} {37Gb/s} {GDDR7 DRAM} with {PAM3}-optimized {TRX} equalization and {ZQ} calibration,'' in \emph{IEEE ISSCC}, 2024, pp. 242--244.

\bibitem{nvidia2023grace}
{NVIDIA Corporation}, \emph{{NVIDIA Grace CPU} Superchip Architecture Whitepaper}, 2023.

\bibitem{park2023lr112g}
H.~Park \emph{et~al.}, ``A {4.63pJ/b} {112Gb/s} {DSP}-based {PAM-4} transceiver for a large-scale switch in 5nm {FinFET},'' in \emph{IEEE ISSCC}, 2023, pp. 112--114.

\bibitem{credo2025weaver}
{Credo Technology Group}, \emph{{OmniConnect Weaver} Memory Fanout Gearbox Product Brief}, 2025.

\bibitem{deepseekv2_2024}
A.~Liu \emph{et~al.}, ``{DeepSeek-V2}: A strong, economical, and efficient mixture-of-experts language model,'' arXiv:2405.04434, 2024.

\bibitem{bai2020sedram}
F.-J. Bai \emph{et~al.}, ``A stacked embedded {DRAM} array for {LPDDR4/4X} using hybrid bonding {3D} integration with 34{GB/s/1Gb} {0.88pJ/b} logic-to-memory interface,'' in \emph{IEEE IEDM}, 2020, pp. 6.6.1--6.6.4.

\bibitem{chen2025hbm3d}
Y.~Chen \emph{et~al.}, ``Breaking thermal bottleneck in {3D} {HBM}-on-{GPU} integration via system-technology co-optimization,'' in \emph{IEEE IEDM}, 2025.

\bibitem{11113276}
G.~Karfakis \emph{et~al.}, ``Optimizing thermal performance in {2.5D} systems using embedded isolators,'' \emph{IEEE J. Emerg. Sel. Topics Circuits Syst.}, vol.~15, no.~3, pp. 458--468, 2025.

\end{thebibliography}
